\documentclass[twocolumn,aps,prl,10pt,amsmath,amssymb,nofootinbib,showpacs,superscriptaddress,floatfix]{revtex4-1}

\newcommand{\rs}{\rm\scriptscriptstyle}

\DeclareFontFamily{U}{rcjhbltx}{}
\DeclareFontShape{U}{rcjhbltx}{m}{n}{<->rcjhbltx}{}
\DeclareSymbolFont{hebrewletters}{U}{rcjhbltx}{m}{n}

\usepackage{graphicx}
\usepackage{color}
\usepackage[usenames,dvipsnames]{xcolor}
\usepackage[colorlinks=true,linkcolor=Red,citecolor=Green,linktoc=page]{hyperref}
\usepackage{multirow}
\usepackage{float}
\usepackage{flushend}
\usepackage{balance}
\usepackage[varg]{txfonts}
\usepackage{ulem}
\usepackage{fancyhdr}

\DeclareMathSymbol{\lamed}{\mathord}{hebrewletters}{108}

\begin{document}
\title{Bose metals, from prediction to realization}

\begin{abstract}
\end{abstract}

\author{M.\,C.\,Diamantini}

\affiliation{NiPS Laboratory, INFN and Dipartimento di Fisica e Geologia, University of Perugia, via A. Pascoli, I-06100 Perugia, Italy}

\author{C.\,A.\,Trugenberger}

\affiliation{SwissScientific Technologies SA, rue du Rhone 59, CH-1204 Geneva, Switzerland}


\
\begin{abstract}
Bose metals are metals made of Cooper pairs, which form at very low temperatures in superconducting films and Josephson junction arrays as an intermediate phase between superconductivity and superinsulation. We predicted the existence of this 2D metallic phase of bosons in the mid 90s, showing that they arise due to topological quantum effects. The observation of Bose metals in perfectly regular Josephson junction arrays fully confirms our prediction and rules out alternative models based on disorder. Here, we review the basic mechanism leading to Bose metals. The key points are that the relevant vortices in granular superconductors are core-less, mobile XY vortices which can tunnel through the system due to quantum phase slips, that there is no charge-phase commutation relation preventing such vortices to be simultaneously out of condensate with charges, and that out-of-condensate charges and vortices are subject to topological mutual statistics interactions, a quantum effect that dominates at low temperatures. These repulsive mutual statistics interactions are sufficient to increase the energy of the Cooper pairs and lift them out of condensate. The result is a topological ground state in which charge conduction along edges and vortex movement across them organize themselves so as to generate the observed metallic saturation at low temperatures. This state is known today as a bosonic topological insulator. 
\end{abstract}
\maketitle

\section{Introduction}
Conventional wisdom has it that metals cannot exist in 1D and 2D \cite{anderson} (for a review see \cite{abrahams}, let alone metals made of bosons. However, already in the first studies of the superconductor-to-insulator transition (SIT) in thin films \cite{goldman1, haviland, hebard} (for a review see \cite{goldmanrev, feigelman}), it was noted that the resistance at the superconductivity-destroying transition levels off at the universal sheet resistance quantum $R_{\rm Q}= h/4e^2 = 6.45 \ k\Omega$. Soon after, Fisher \cite{fisher1, fisher2} showed that bosonic Cooper pairs on the brink of superconductivity can indeed diffuse with the universal quantum of sheet resistance. Finally, we predicted that this metallic quantum critical point can develop to a full intermediate metallic phase as a function of a previously neglected additional quantum parameter \cite{dst}. This intermediate metallic state was experimentally detected the same year \cite{van}. The name ``Bose metal" was adopted in subsequent studies \cite{das, mason} and stuck, together with the variant ``anomalous metal". Actually, the name ``failed superconductor" is often used for the region with saturated $T=0$ sheet resistance $R_{\square} < R_{\rm Q}$ while the name ``failed insulator" is applied to the dual region with $R_{\square} > R_{\rm Q}$. This is misleading, though, since both regions are part of the same intermediate quantum phase, as is now recognized by leading experimental groups, who have re-derived from experimental observations exactly the phase diagram we predicted on theoretical grounds 27 years ago \cite{kapitulniknew}. 

The existence of the Bose metal as a new phase of matter has been put in doubt by claims that it is due only to external perturbations, such as Cooper pairs failing to cool sufficiently or external high-frequency electromagnetic radiation disrupting the measurements \cite{shahar}. Recent experiments, however, have dispelled these notions: a refrigeration problem would be incompatible with the observed resistance saturation onset depending on magnetic field \cite{mason, bm}, while careful filtering of external radiation has confirmed the Bose metal as a new phase of matter \cite{yang, bm} (for a review of these points see \cite{phillipsrev1, phillipsrev2}). 

Since the very first experiments \cite{goldman1}, it was realized that superconducting films near the SIT decompose into emergent granular structures of condensate islands \cite{sacepe1, sacepe2, granular}, with global superconductivity arising from tunnelling in a phase coherent state, best modelled by a Josephson junction array (JJA) \cite{fazio} (for a review see \cite{jjarev}) with random Josephson couplings and charging energies, this fragmentation being caused by strong infrared divergences in the 2D limit \cite{planar}. Superconductivity, thus, can be destroyed either by phase decoherence or by Cooper pairs breaking up. This has lead to models in which the anomalous intermediate metal is of fermionic character. These models have now also been disproved by the results of \cite{yang}, in which the resistance oscillations as a function of applied magnetic field have been measured. The observed periodicity $h/2e$ (instead of $h/e$) proves that the charge carriers are Cooper pairs and the Bose metal is indeed bosonic. 

The perceived problem with a metal made of bosons arises from the commutation relation
\begin{equation}
\left[ Q, \varphi \right] = i \ ,
\label{cp}
\end{equation}
between particle number and phase, which would imply that either the phase is a sharp observable, with the charge in a condensate or the other way around, 
the charge (number) annihilates the ground state and the global U(1) symmetry is unitarily realized, with phases in a random state, which is typically considered a vortex Bose condensate \cite{fisher1}. Superconductor and Mott insulator due to vortex condensation seem to be the only possible phases of bosons at $T=0$ see, e.g. \cite{phillips3, phillips4}. Unfortunately, however, this conclusion is erroneous, since a phase operator does not exist due to the periodicity of the angle (for a review see \cite{carruthers,  karstrup}); a global phase operator may only be defined in a superfluid (superconducting) state \cite{rocca}. As is well-known from all models with compact U(1) symmetries, one must first correctly identify and separate the topological excitations due to compactness; only then can one promote the original angles to real variables on the infinite line, thereby eliminating the problems in commutators like (\ref{cp}). We will show that, indeed, there is nothing preventing, in principle, a ground state with out-of-condensate charges and random phases but with no vortex condensate. 

One way to obtain such a state is to invoke disorder, in form of statistically distributed Josephson couplings, leading to a picture of the Bose metal as a phase glass \cite{phillips3}. A first hint that disorder is not relevant, however is already provided by the experiments \cite{clean1, clean2}, in which the Bose metal is observed in clean crystalline samples. The final blow for disorder-based models, however is the detection of the Bose metal in perfectly regular, disorder-free JJAs \cite{van, marcus1, marcus2, marcus3}. If Bose metals are both predicted and confirmed to arise in perfectly clean, regular systems, any additional disorder cannot be instrumental to their formation. These JJA experiments, of course, rule out also models predicated on dynamical dimensional reduction \cite{geom}. This is not a problem, though; the original simple explanation we gave when we predicted the Bose metal does indeed apply also to regular JJAs: the global U(1) symmetry is indeed unbroken except on the edges of the system (for a review see \cite{book}) and the Bose metal is indeed a Mott insulator, albeit a topological one, save on the (possibly also internal) edges, where the metallic conduction comes from. The idea of a frozen phase state is essentially correct; however, as is well known, disorder is not the only mechanism to obtain frozen states, frustration is another one. When both charges and vortices are out-of-condensate and can tunnel on a JJA and its dual lattice, motion of vortices would cause an energy increase for charges in the orthogonal direction and vice versa. The system is frustrated and nothing can move without increasing the energy, so that the system is trapped in one of many local minima. The only locations where this frustration can be overcome is at the edges. The resulting frozen state has nothing to do with exogeneous disorder \cite{phillips3}. The origin of this state lies ultimately in the frustration caused by the mutual statistical interactions between charges and vortices, which are relative fermions, and is present also in perfectly ordered systems \cite{van, marcus1, marcus2, marcus3}. In the rest of the paper we will explain in detail how this frustrated state forms. 

Let us first focus on the properties of the vortices near the SIT. Because of the JJA-like granular structure of the condensate in this region, the relevant vortices are not intra-grain Abrikosov vortices (for a review see \cite{tinkham}), but, rather, the inter-grain XY-like vortices (for a review see \cite{minnhagen}) due to non-trivial circulations of the phases on neighbouring condensate islands. Such vortices are point-like structures with no dissipative core. As a consequence they are extremely mobile because of quantum phase slips \cite{golubev} (for a review see \cite{arutyunov}). 

\begin{figure}[t!]
	\includegraphics[width=9cm]{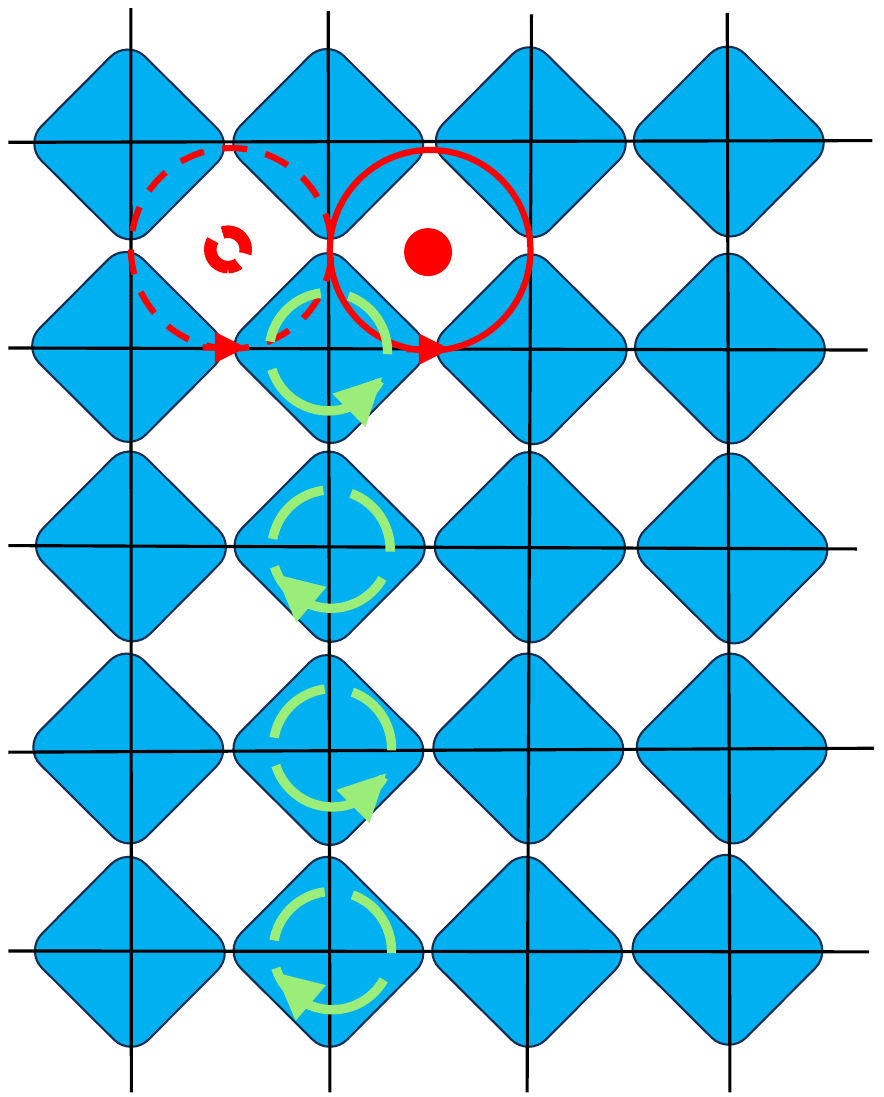}
	\vspace{-0.3cm}
	\caption{A phase slip on a 2D JJA, corresponding to vortex tunnelling on the dual lattice. }
	\label{Fig1}
\end{figure}

Quantum phase slips (QPS) are tunnelling events in which the phase on one condensate island makes a flip by an angle $\pm 2\pi$ to return in the same position, corresponding to what is known as a large gauge transformation. For some reason, QPS are prominently discussed as fundamental aspects for the low-temperature properties of 1D Josephson junction chains \cite{golubev, arutyunov} but are rarely taken into account for JJA and films in 2D. Here too, though they are of paramount importance. Because of the U(1) gauge symmetry, the only gauge invariant degrees of freedom encoded in the grain phases are the vortices living on the plaquettes of the array, i.e. on the dual lattice. Let us consider an infinite line of phase slips of alternating signs ending on a particular site. This is everywhere a pure gauge configuration except on the end site of the line, where the slip corresponds to a vortex tunnelling from one plaquette to an adjacent one, as shown in Fig.\,\ref{Fig1}. 

Because of phase slips, the inter-grain XY vortices tunnel on the dual lattice, exactly as Cooper pairs tunnel on the main grain lattice due to the Josephson effect. To fully describe the physics of JJAs and thin superconducting films near the SIT one {\it must} add a tunnelling term for vortices dual to the Cooper pair tunnelling term, as we pointed out originally in \cite{dst}. As always when kinetic terms are added to an action, this alters the canonical structure of the model. As a consequence, the absence of whatever obstruction to forming a Bose metal phase becomes even more evident. In a nutshell, QPS and the vortex motion hamper charge transport by generating a long-range Coulomb potential along its trajectory while, vice versa, charge Josephson currents hamper vortex motion by the dual mechanism. The result is a frozen ground state in which both charges (in all what follows, by charges we always mean bosonic Cooper pairs) and vortices are out of condensate and in which charge diffusion survives along percolation lines with vortex diffusion across them leading to the measured metallic resistance. 

In this state, the relative statistical interaction deeply couples charges and vortices. A vortex is always accompanied by circular charge currents around it, which cost an energy. These, on the other side, create Coulomb energies across them, further increasing the energy cost to create a vortex. The dual picture holds, of course, for charges. Although there is no condensate, there is a gap for both charges and vortices. 

We now explain formally how this comes about. To this end let us turn the problem around and ask how the quantum state of a system of out-of-condensate charges and vortices must look like. We consider thus $N_{\rm e}$ electric charges $Q_I$ (integer units of $2e$) labelled by the integers $I=1\dots N_{\rm e}$ and $N_{\rm m}$ magnetic vortices $M_J$ (integer units of $\Phi_0= 2\pi/2e$) labelled by the integers $J=1\dots N_{\rm m}$ (henceforth we use natural units $c=1$, $\hbar = 1$, $\varepsilon_0=1$ for simplicity of presentation). The main characteristic of such a system is that the wave function of each type of excitation acquires a topological phase when circling around the other type of excitation. These are the famed Aharonov-Bohm \cite{aharonovbohm} and Aharonov-Casher \cite{aharonovcasher} phases and constitute the simplest example of what is known as relative statistics phases \cite{wilczek}: charges and vortices are relative fermions. 

Following \cite{wu}, the Euclidean partition function of the system of charges and vortices in the path integral formulation (see e.g. \cite{negele}) is given by
\begin{equation}
Z = \int {\cal D} Q {\cal D} M \ {\rm e}^{-S_0+i S_{\rm Top}} \ ,
\label{wuform}
\end{equation}
where ${\cal D} Q = {\cal D} x_1(s) \dots {\cal D} x_{N_e} (s)$, with an analogous expression for ${\cal D}M$, and $x_I(s)$ and $x_J(s)$ parametrize the charge and vortex trajectories $C_I$ and $C_J$, respectively. In this expression, $S_0$ is the action, comprising the usual kinetic and potential energy terms,
while the phase encodes the relative statistics phases and is given by
\begin{eqnarray}
S_{\rm top} &&=  2\pi  \sum_{I,J} Q_I M_J \Phi \left( C_I, C_J \right) \ ,
\nonumber \\
\Phi \left( C_I, C_J \right) &&= {1\over 4\pi} \int_0^1 ds \int_0^1 dt \ {dx_I^{\mu} \over ds} \epsilon_{\mu \alpha \nu}
{(x_I -x_J)^{\alpha} \over |x_I-x_J|^3} {dx_J^{\nu} \over dt} \ .
\label{gauss}
\end{eqnarray} 
For closed curves $C_I$ and $C_J$, the quantity $\Phi \left( C_I, C_J \right)$ represents the integer Gauss linking number, the simplest topological knot invariant, see \cite{kaufmann}. If the excitations satisfy the quantization condition $Q_I M_J = {\rm integer} $ for all $I, J$, the two species are relative fermions and the above phases become trivial for closed trajectories, corresponding, e.g., to a Minkowski space-time fluctuation creating a charge and a hole that annihilate after having encircled a vortex. In general, however, relative statistics phases lead to non-trivial quantum interference effects that cannot be neglected. 

Unfortunately, the phases in (\ref{gauss}) are non-local, and thus unsuited for the machinery of quantum field theory. This can be remedied, however, by the introduction of emergent gauge fields \cite{wilczek}. To see how this works we first introduce introduce the charge and vortex current fields
\begin{eqnarray}
Q_{\mu} &&= \sum_I Q_I \int ds \ {dx_I^{\mu} \over ds} \ \delta^3 \left( x -x_I (s) \right) \ ,
\nonumber \\
M_{\mu} &&= \sum_J M_J \int dt \ {dx_J^{\mu} \over dt} \ \delta^3 \left( x -x_J(t) \right) \ .
\label{pointsources}
\end{eqnarray}
Then we can represent the relative statistics phases as 
\begin{equation}
S_{\rm top}  =  \ 2\pi   \int d^3x \  Q_{\mu} \epsilon_{\mu \alpha \nu} {\partial_{\alpha} \over \nabla^2} M_{\nu} \ ,
\label{effec3}
\end{equation}
where $\epsilon^{\mu \alpha \nu}$ denotes the totally antisymmetric tensor and 
\begin{equation}
{1\over -\nabla^2} \delta^3 (x) = {1\over 4\pi} {1\over |x|} \ ,
\label{3dG}
\end{equation} 
is the 3D Poisson Green function. 

This is still non-local. One can, however obtain a local formulation \cite{wilczek} by coupling the charge and vortex currents to two gauge fields $a_{\mu}$ and $b_{\mu}$ with a mixed Chern-Simons interaction \cite{jackiw1, jackiw2},
\begin{eqnarray}
Z_{\rm top} &&= \int {\cal D}Q_\mu {\cal D} M_\mu  \int {\cal D} a_{\mu} {\cal D} b_{\mu} {\rm e}^{-S} \ ,
\nonumber \\
S_{\rm top} &&= \int d^3 x {i\over 2\pi} a_{\mu} \epsilon^{\mu \alpha \nu} \partial_{\alpha} b_{\nu} + i a_{\mu} Q_{\mu} + i b_{\mu} M_{\mu} \ .
\label{wilc}
\end{eqnarray}
This is clearly a local action, involving only first derivatives. When integrating over the two emergent gauge fields, though we obtain back the non-local formulation (\ref{effec3}). Note that both the parity and time-reversal symmetries are respected since $a_{\mu }$ is a vector field, as usual, but $b_{\mu}$, coupling to the vortex current, is a pseudovector gauge field. In modern parlance, this action, first derived as an intermediate state of the SIT in \cite{dst} describes a bosonic topological insulator \cite{bti1, bti2, bti3}. Since it is purely topological, without any dynamics \cite{jackiwbook}, there are no excitations and the equations of motion, derived from the Minkowksi version of (\ref{wilc}) are exact: 
\begin{eqnarray}
f^{\mu} = {1\over 2} \epsilon^{\mu \alpha \nu}  f_{\alpha \nu} = \epsilon^{\mu \alpha \nu} \partial_{\alpha} b_{\nu} =2\pi Q^{\mu} \ ,
\nonumber \\
g^{\mu} = {1\over 2} \epsilon^{\mu \alpha \nu}  g_{\alpha \nu} = \epsilon^{\mu \alpha \nu} \partial_{\alpha} a_{\nu} =2\pi M^{\mu} \ .
\label{fields}
\end{eqnarray}
They imply that the dual field strengths $f^{\mu}$ and $g^{\mu}$ associated with the two emergent gauge fields represent ($2\pi$ times) the conserved charge and vortex (number) currents, respectively. 

There is an alternative point of view, though \cite{dst}. In the discrete formulation, the integers $Q_{\mu}$ and $M_{\mu}$ represent the topological excitations that render the dual field strengths $f_{\mu}$ and $g_{\mu}$ periodic in a Villain-type formulation and implement, thus the compactness of the two U(1) gauge symmetries. When these topological excitations are explicitly taken into account, as in (\ref{wilc}), they can be reabsorbed into the gauge fields rendering both these and the corresponding gauge functions real variables defined on the infinite line. One can then proceed to canonically quantize the remaining double Chern-Simons theory. 

Since the action is of first-order in time derivatives, the canonical commutation relations can be simply read off the action after setting the Lagrange multipliers $a_0=0$ and $b_0=0$, the Weyl gauge for the gauge theory \cite{djt}, 
\begin{equation}
\left[ a_i({\bf x}), b_j ({\bf y}) \right] = -i \ 2\pi \ \epsilon^{ij} \delta^2 \left({\bf x} -{\bf y} \right)\ ,
\label{comm1}
\end{equation}
where boldface characters denote purely spatial coordinates. If we decompose the spatial gauge fields into longitudinal and transverse components, $a_i = \partial_i \varphi + \epsilon^{ij} \partial_j \chi $ and $b_i = \partial_i \theta + \epsilon^{ij} \partial_j \xi$ these commutation relations imply
\begin{eqnarray}
\left[ Q, \varphi \right] &&= i \ ,
\nonumber \\
\left[ M, \theta \right] &&=i \ ,
\label{comm2}
\end{eqnarray}
where $Q=Q^0$ and $M=M^0$ are the particle- and vortex number operators, respectively. Now, however, the variables $\varphi$ and $\theta$ are real variables defined on the infinite line and there is no problem with the definition of these commutators. When vortices are correctly treated as fundamental excitations, which can tunnel through the system, exactly as Cooper pairs, two global U(1) symmetries are present, one generated by the particle number and one by the vortex number. There is absolutely no obstruction to both generators annihilating the ground state and both symmetries being unitarily implemented so that neither charges nor vortices are in a condensate. 

But, if the ground state is not a condensate and there are no low-energy bulk excitations, how can it conduct at $T=0 \ ^\circ K$? The global U(1) symmetry still seems to forbid this. 
The crucial point is that there are gapless edge excitations, and it is these that conduct at $T=0$. Indeed, the Chern-Simons action (\ref{wilc}) is gauge invariant only in the bulk of the system, not on its boundaries \cite{jackiwbook}. Exactly as for the quantum Hall states \cite{wen}, there are massless edge excitations, in this case electric ones propagating along the edges and magnetic ones crossing the edges. From the point of view of the edges, these correspond to 1D quantum phase slips, resulting in a finite metallic resistance \cite{bm, book}. 

To derive the boundary physics let us minimally couple the Chern-Simons action to the real electromagnetic field $A_{\mu}$ (in Minkowski space-time),
\begin{equation}
S={1\over 2\pi} \int d^3x \ a_{\mu} \epsilon^{\mu \alpha \nu} \partial_{\alpha} b_{\nu} - {2e\over 2\pi} \int d^3x \ A_{\mu} \epsilon^{\mu \alpha \nu} \partial_{\alpha} b_{\nu}. 
\label{mincou}
\end{equation}
This action is not gauge invariant under transformations that do not vanish on the boundary and one must add edge degrees of freedom so that the gauge invariance of the complete model is restored, exactly as for the quantum Hall states \cite{wen}. To do so we consider the Weyl gauge $a_0=0$, $b_0=0$ and we choose the spatial field components as pure gauges, $a_i=\partial_i \lambda$, $b_i=\partial_i \chi$. To restore full gauge invariance, these fields have to be promoted to new scalar degrees of freedom on the edges, with an edge action
\begin{equation}
S_{\rm e} = {1 \over 4\pi} \int d^2 x \ \left( \partial_0 \lambda \partial_s \chi + \partial_0 \chi \partial_s \lambda \right) +
{2e \over 2\pi} \int d^2 x \  \left( A_0  \partial_s \chi - A_s \partial_0 \chi \right) ,
\label{edgeac}
\end{equation}
where $s$ denotes the space coordinate along the edge. We now introduce two new edge fields by the relations $\lambda = \xi + \eta$ and $\chi= \xi-\eta$ and we set $A_s=0$ to focus on a configuration with a constant edge electric field $E=\partial_s A_0$,
\begin{equation}
S_{\rm e} = {1 \over 2\pi} \int d^2 x \ \left( \partial_0 \xi \partial_s \xi - \partial_0 \eta \partial_s \eta \right) +
2e  \int d^2 x \   A_0  \left( {1\over 2\pi} \partial_s \chi \right) .
\label{edgeacmod}
\end{equation}
From here we identify $\rho = \partial_s \chi /2\pi$ as the one-dimensional charge density of edge excitations. 

As in the quantum Hall case \cite{wen}, we must add a non-universal dynamics for the edge modes, embodied in the Hamiltonian
\begin{equation}
H_{\rm e} = {1 \over 2\pi} \int ds \left[ v_b \left( \partial_s \xi \right)^2 + v_b \left( \partial_s \eta \right)^2\right] \ ,
\label{four}
\end{equation}
where $v_b$ is the velocity of propagation of the edge modes along the boundary. Upon adding this term, the total edge action becomes
\begin{eqnarray}
S_{\rm e} &&= {1 \over 2\pi} \int d^2 x \ \left[ \left(\partial_0-v_b\partial_s\right) \xi \partial_s \xi  - \left(\partial_0+v_b \partial_s \right)\eta \partial_s \eta \right]\nonumber \\
&&+ 2e  \int d^2 x \  A_0  \left( {1\over 2\pi} \partial_s \chi \right) \ ,
\label{edgeacfull}
\end{eqnarray}
with the corresponding equations of motion
\begin{eqnarray} 
\left( \partial_0 -v_b \partial_s \right) \partial_s \xi &&= - e E \ ,
\nonumber \\
\left( \partial_0 +v_b \partial_s \right) \partial_s \eta &&= - e E \ ,
\label{eq1}
\end{eqnarray}
or, in the original fields,
\begin{eqnarray} 
\partial_0 \partial_s \lambda -v_b \partial_s^2 \chi &&= - 2e E \ ,
\nonumber \\
\partial_0 \partial_s \chi - v_b \partial_s^2\lambda &&= 0\ .
\label{eq2}
\end{eqnarray}
Using the edge current $I = 2e v_b \rho$, we can rewrite the first of these equations as
\begin{equation}
\partial_s I = {4e^2 \over 2\pi} E + {2e\over 2\pi} \partial_0 \partial_s \lambda \ .
\label{fincur}
\end{equation}
Remembering that $(1/2\pi) \epsilon^{\mu \alpha \nu} \partial_{\alpha} a_{\nu}$ is the vortex number current, we note that $(1/2\pi) \partial_0 \partial_s \lambda$ is the number of vortices crossing the edge in a given orthogonal direction. On the right-hand side of (\ref{fincur}) we see the two mechanisms by which charges are accelerated/decelerated along the edge, an external electric field and QPS. 

To compute the resistance of the edges we will time-average the above equation for the current, to obtain the steady-state value, 
\begin{equation}
\langle I \rangle = {4e^2\over 2\pi} \left( V + {1\over 2e} \langle \partial_0 \lambda \rangle \right) \ ,
\label{add1}
\end{equation}
where $V$ is the applied voltage, $E= \partial_s V$. This shows that QPS act as an additional, effective voltage on the edge. If we define this as
\begin{equation}
V_{\rm QPS} = {1\over 2e} \langle \partial_0 \lambda \rangle = \kappa V \ ,
\label{add2}
\end{equation}
we can express the induced current as
\begin{equation}
\langle I \rangle = {4e^2\over 2\pi} (1+\kappa) V \ ,
\label{add3}
\end{equation} 
which gives the sheet resistance
\begin{equation}
R_{\square} = {2\pi \over g (2e)^2} = {R_{\rm Q} \over g} 
\label{qr}
\end{equation}
where $g=(1+\kappa)$. This can be viewed as the resistance quantum with an effective electron charge $e_{\rm eff} =e \sqrt{g}$ renormalized by the QPS interactions. 

At the self-dual point, the ``collisions" with QPS are completely random and as likely to accelerate or decelerate an edge Cooper pair. As a result, $\kappa=0$ because the time-average of $\partial_0 \lambda$ vanishes and we obtain exactly the resistance quantum. On the superconducting side, there is an overabundance of accelerating QPS ``collisions" and the resulting resistance is smaller than the quantum. On the contrary, on the insulating side, there is an overabundance of decelerating ``collisions" and the resulting resistance is larger than the quantum. 

Two points are important to mention about this result. First of all, the edges on which there is metallic charge transport need not be the external edges of the sample; there can be as well internal edges forming a percolation network, exactly as in the quantum Hall states \cite{cc}. Second, edge charge transport is symmetry-protected from localization and disorder-induced dissipation by its correspondence with a topological bulk ground state \cite{bti1, bti2, bti3}: quantum phase slips are the only dissipation mechanism. It is now established experimentally that vortices are unpaired and mostly frozen in the Bose metal state while a fraction remain mobile and cause the residual metallic sheet resistance saturation \cite{marcus3}. 

We have derived what is the origin of the longitudinal metallic resistance distributed around the quantum. Let us now focus on bulk physics. Having established that gauge fields are necessary for a local description of the Bose metal, we can now proceed to construct the complete effective field theory of the system, as usual, by adding the most general gauge invariant terms in a derivative expansion and retaining only power-counting relevant and marginal terms. The next order terms in the expansion of the Euclidean action, involving two derivatives of the fields, are the usual Maxwell terms for the two gauge fields, 
\begin{equation}
S=\int d^3 x \  {i\over2 \pi} a_{\mu} \epsilon^{\mu \alpha \nu} \partial_{\alpha} b_{\nu} 
+{1\over 2e^2_v} f_{\mu}f_{\mu} 
+{1 \over 2e^2_q} g_{\mu} g_{\mu} 
+i a_{\mu}  Q_{\mu} +i b_{\mu}  M_{\mu} \ . 
\label{nonrelac3}
\end{equation}
The two parameters $e_{\mathrm q}^2$ and $e_{\mathrm v}^2$ represent the orders of magnitude of the electric and magnetic energies of an elementary electric grain of charge $2e$ and size $d$ and an elementary magnetic grain with flux quantum $\Phi_0=\pi/e$ and size $\lambda_\perp = \lambda_{\rm L}^2/d$, the Pearl length, 
\begin{eqnarray}
e_{\mathrm q}^2 &&= {\cal O} \left( {4e^2\over  d} \right) \ ,
\nonumber \\
e_{\mathrm v}^2 &&={\cal O} \left(  {\Phi_0^2 \over \lambda_{\perp}}\right)  = {\cal O} \left( {\pi^2 \over e^2 \lambda_{\perp} } \right) = {\cal O} \left( {\pi^2 d \over e^2 \lambda_{\rs L}^2}\right)  \ .
\label{couplings}
\end{eqnarray}
They represent the two typical energy scales in the problem, and their ratio determines the relative strength of magnetic and electric forces. For ease of presentation we have chosen relativistic notation. Non-relativistic effects can be easily incorporated and are discussed extensively in \cite{book}. 
The two parameters $e_{\mathrm q}^2$ and $e_{\mathrm v}^2$ can be traded for one massive parameter, the topological Chern-Simons mass \cite{jackiw1, jackiw2} $m=e_{\mathrm q}e_{\mathrm v}/2\pi $ and one dimensionless parameter $g=e_{\mathrm v}/e_{\mathrm q} = O(d/(e^2 \lambda_L))$.

\begin{figure}[t!]
	\includegraphics[width=9cm]{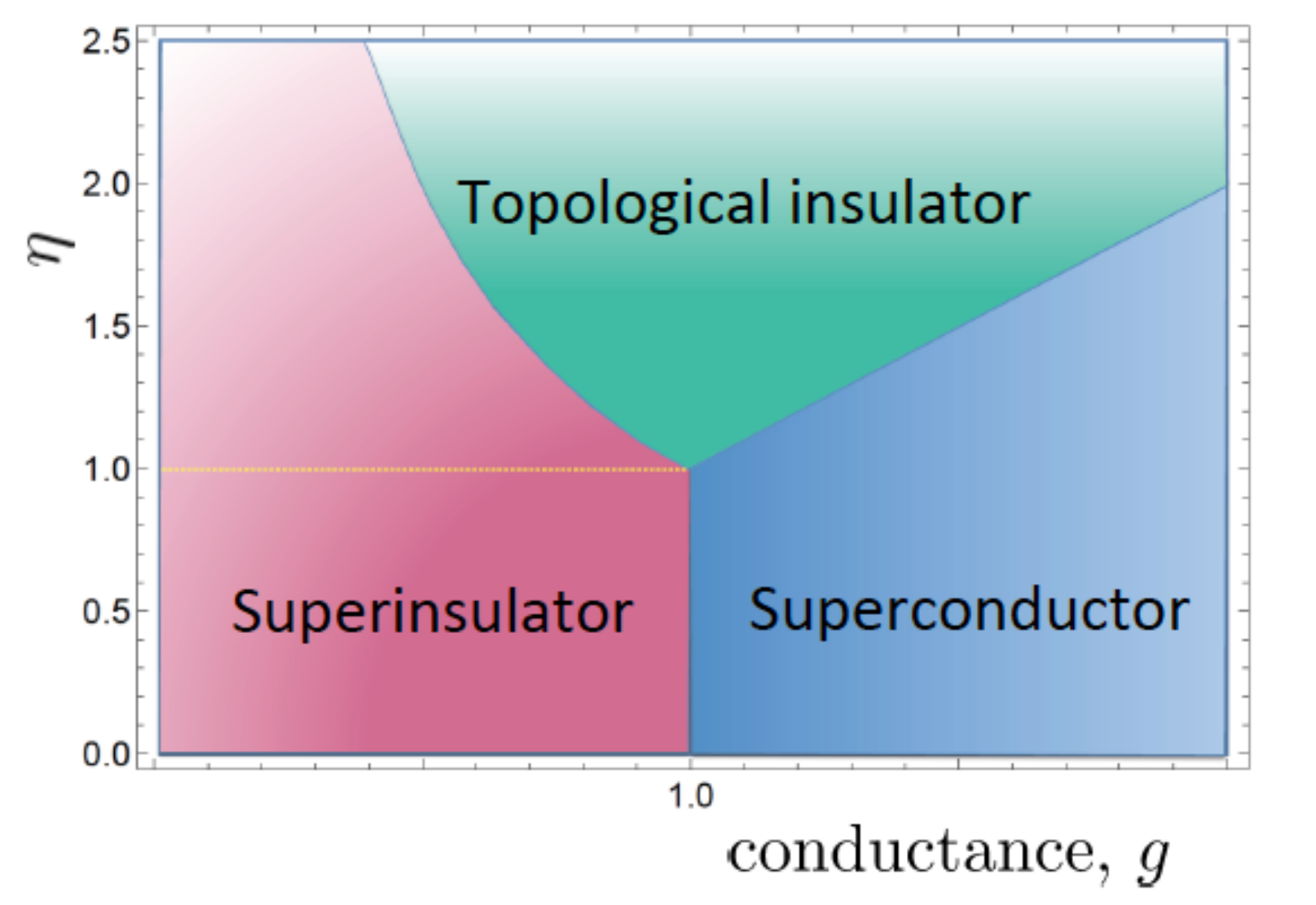}
	\vspace{-0.3cm}
	\caption{The quantum phase structure of JJA and superconducting films as a function of the two parameters $g$ and $\eta = f(m\ell)$. The bosonic topological insulator phase is also called Bose metal due do its edge metallic behaviour. }
	\label{Fig2}
\end{figure}

Given that $e_{\mathrm q}^2$ and $e_{\mathrm v}^2$ have canonical dimension [mass] and, correspondingly, the two kinetic terms in (\ref{nonrelac3}) are infrared-irrelevant, one might be tempted to leave them out in the effective field theory construction and consider anyway only the infrared-dominant mixed Chern-Simons theory, obtained in the limit $e_{\mathrm q}^2 \to \infty$ and $e_{\mathrm v}^2 \to \infty$. The problem with this is that this limit is not well defined without specifying the value of $g$ in the limit. We must thus keep both kinetic terms in the derivative expansion of the effective action, even if each is individually infrared-irrelevant. But then, of course we realize immediately that we have a second, additional dimensionless parameter in the game: this is the ratio of the intergrain distance $\ell$ to the topological length scale $1/m$, the parameter $m\ell$. The behavior of charges and vortices depends crucially on both dimensionless parameters and, as a consequence, one obtains very different ground states when they are varied. The quantum phase structure has been derived in \cite{dst} (for a review see \cite{book}) as a function of $g$ and $\eta$, which is a function of $m\ell$ and is shown in Fig.\,\ref{Fig2}. This is exactly the now experimentally confirmed phase structure of JJAs \cite{marcus3} and superconducting films \cite{bm, kapitulniknew}. 

The two quantum Berezinskii-Kosterlitz-Thouless (BKT) transitions \cite{ber, kt1, kt2} to the superconducting and superinsulating states take place at the dual points $g=\eta >1$ and $g=1/\eta <1$, another experimentally verified fact \cite{bm, book}. Finally, we have identified the charge renormalization factor $g$, giving rise to the metallic resistance regime (\ref{qr}), with one of the two quantum parameters driving the SIT. This identification is strongly suggested by experiments \cite{bm, book}. 

Note that, when adding the required kinetic terms (\ref{nonrelac3}), the canonical structure of the theory is changed, since we have now terms quadratic in time derivatives in the action. Correspondingly, the commutation relations (\ref{comm2}) are changed to
\begin{eqnarray}
\left[ Q +{1\over 2e_q^2} \dot \varphi, \varphi \right] &&= i \ ,
\nonumber \\
\left[ M+{1\over 2e_v^2} \dot \theta, \theta \right] &&=i \ .
\label{comm3}
\end{eqnarray}

As we now show, the intermediate Bose metal phase is a longitudinal analogue of the quantum Hall plateaus, although, in this case, we do not have a real ``plateau".  To this end let us focus on the microscopic picture. As we mentioned previously, charges and vortices are mutual fermions \cite{wilczek} and are therefore frozen into a bosonic topological insulator phase when their effective masses are large enough to prevent them from condensing. The wave function for such a ground state can be explicitly constructed \cite{nodisorder}. We consider first charges and vortices of one single chirality and we 
adopt the usual mean field approximation of anyon physics (for a review see \cite{wilczekbook}) by replacing the homogeneously distributed vortices by the uniform magnetic field 
\begin{equation}
{\cal B}_{\Phi} = {2e N_{\Phi} \Phi_0 \over A } \ ,
\label{magnetic}
\end{equation}
felt by the Cooper pairs, where $N_{\Phi}$ is the number of vortices and $A$ the area of the sample, and the charge distribution with the corresponding uniform ``magnetic field" 
\begin{equation}
{\cal B}_{Q} = {\Phi_0 N_Q 2e \over A }\ ,
\label{magnus}
\end{equation}
felt by the vortices via the Magnus force, the dual of the Lorentz force \cite{magnus}. The hydrodynamic Magnus force is experienced by a vortex moving with the velocity  ${\bf v}$ over the charge background, and in 2D it becomes 
$F_{\rm M}^i = \kappa \epsilon^{ij} v^j \rho$ where $\rho$ is the fluid density and $\kappa$ is the circulation of the vortex expressed in multiples of $2\pi$. When the fluid is charged, in units of $2e$, it can be written as $F_{\rm M}^i = \Phi  \epsilon^{ij} v^j 2e \rho $
where $\Phi$ is the total vorticity in multiples of $\Phi_0$. This is the exact dual of the usual Lorentz force $F_{\rm L}^i =
2e \epsilon^{ij} v_j B$, from which we evince that the charge density plays for vortices exactly the same role as the magnetic field for charges. 

At the self-dual point, where $N_{\mathrm Q} = N_\Phi$, the two ``magnetic fields" ${\cal B}_{\mathrm Q}$ and ${\cal B}_\Phi$ become identical,
${\cal B}_{\mathrm Q} ={\cal B}_\Phi = {\cal B}$ and we can use the known integer Hall ground state for two flavours of bosons in a magnetic field \cite{senthillevin}, 
\begin{equation}
	\Psi \left( \{ z_i\} , \{ w_i \} \right) = \prod_{i<j} |z_i-z_j| \prod_{i<j}|w_i-w_j| 
	\prod_{i,j} {(z_i-w_j )\over |z_i-w_j|} \ {\rm e}^{ -{{\cal B} \over 4} \left( \sum_i |z_i|^2 + |w_i|^2 \right) } \ ,
	\label{special}
\end{equation}
where $z_i$ and $w_j$ denote the positions of the two flavours of particles on the complex plane. We can now repeat the exact same construction for vortices of the opposite chirality, corresponding to a mean field magnetic field of the opposite sign. This will give the corresponding  wave function of the opposite chirality, 
\begin{equation}
	\Psi \left( \{ \bar z_i\} , \{ \bar w_i \} \right) = \prod_{i<j} |\bar z_i-\bar z_j| \prod_{i<j}|\bar w_i-\bar w_j| 
	\prod_{i,j} {(\bar z_i-\bar w_j )\over |\bar z_i-\bar w_j|} \ {\rm e}^{ -{{\cal B} \over 4} \left( \sum_i |z_i|^2 + |w_i|^2 \right) } \ ,
	\label{special}
\end{equation}
Finally, we use the fact that, in the topological insulator state, vortices acquire a minus sign in their wave function when changing direction under time reversal, contrary to the trivial insulator state \cite{xu}. A candidate state for the topological insulator is thus
\begin{equation}
\Psi \left( \{ z_i\} , \{ w_i \} \right) + {\rm e}^{-i{N_{\Phi }\pi}} \Psi \left( \{ \bar z_i\} , \{ \bar w_i \} \right) \ .
\label{topoinsu}
\end{equation}

Exactly as in the analogous quantum Hall effect situation, small deviations from the self-dual point lead to small excess densities of quasi-particle and quasi-hole excitations of the two kinds. It comes as no surprise that, for an integer topological state, these excitations are charges and vortices themselves\,\cite{wilczek}. As recently pointed out in \cite{kivelson}, for sufficiently small densities near the self-dual point, these excitations out of the topological state behave as classical particles with fixed guiding-centers in the respective magnetic fields. Since both have long-range Coulomb interactions, they will form a Wigner crystal below a certain critical temperature. Moreover, since the effective magnetic fields involved are very large, this critical temperature is quite high \cite{kivelson}. While excess particles are pinned in the Wigner crystal configuration, the conductivity is due exclusively to the topological state. 

Now comes the crucial difference with respect to the familiar quantum Hall case. The magnetic field is not an external field but it is a dynamical field generated by one flavour of particle for the other. When we have excess particles pinned in the Wigner crystal, the remaining particles available to the topological state change from the original values to new values $N_{\rm Q} \to N_{\rm Q}^{\prime}$ and $N_\Phi \to N_\Phi^{\prime}$. In order to maintain the balance of the two mutual mean field magnetic fields, the system must react by ``dressing" charges and vortices so that they have effective charge $e^{\prime}$ and vorticity $\Phi_0^{\prime}$ satisfying 
\begin{equation}
	e^{\prime} N_{\Phi}^{\prime} \Phi_0 = e N_{\mathrm Q}^{\prime} \Phi_0^{\prime} \ ,
	\label{balance1} 
\end{equation}
Finally, if we require that the integer mutual statistics is maintained we will have again an integer topological quantum state but with a different, renormalized coupling constant
\begin{equation}
	e \to e^{\prime} = e\sqrt{g}\,, \,\,\,\,\,\,\,
	g ={N_{\mathrm Q}^{\prime} \over N_{\Phi}^{\prime} } \ ,
	\label{renor}
\end{equation}
which leads to a renormalized resistance (\ref{qr}). Here we have yet another interpretation of the SIT quantum parameter $g$ in the Bose metal phase, as the ratio of charges to vortices in the topological state. The Bose metal is thus the longitudinal analogue to a quantum Hall plateau, the difference being that the resistance is not a constant in the ``plateau", since the magnetic field is not external but a dynamical one, determined self-consistently from the state itself. 

We are finally ready to explain the finite temperature behaviour. In the ``failed insulator" region, there are more vortices than charges in the topological state, see (\ref{renor}), which means that there is an excess of charges pinned in the Wigner crystal. In general, the two gaps for the bulk of the topological state and for the Wigner crystal will be different. If the excess charges in the Wigner crystal are activated before than the topological bulk, when raising the temperature, the sheet resistance of the overall system will decrease, as observed. 
There are two parallel channels for this decrease: one is the constant edge channel, the other is the activated bulk. Indeed, it has been experimentally confirmed \cite{bm} that the sheet resistance as a function of temperature fits excellently the formula for two parallel resistors,
\begin{equation}
R_{\square} = { R_{\rm bulk}(T) \ R_{\rm edge} \over R_{\rm bulk}(T) + R_{\rm edge}} \ ,
\label{parallel}
\end{equation}
where $R_{\rm edge}$ is a constant corresponding to the metallic saturation, while $R_{\rm bulk}(T)$ describes an activated behaviour. The ``failed superconductor" is simply the dual state with charges and vortices interchanged. In this case the sheet resistance increases with temperature due to an excess of vortices in the Wigner crystal, which get activated first. 

\section{Conclusions}
Bose metals are frozen states of Cooper pairs and vortices caused by the frustration induced by mutual statistics and are both predicted and realized in clean and regular systems. Like fermions can never condense, mutual fermions cannot condense in an intermediate range where electric and magnetic couplings are comparable. In this out-of-condensate state, the statistical repulsion between Cooper pairs and vortices generates a frozen state in which bulk excitations have a large gap. Conduction is realized by gapless excitations, the metallic resistance arising from quantum phase slips due to vortices crossing the edges.

\end{document}